\documentclass[12pt,titlepage]{amsart}
\usepackage{graphicx}
\usepackage{amssymb}
\usepackage{lscape}
\usepackage{epsfig}
\usepackage{bm}
\newcommand{\Mod}[1]{(\!\!\!\!\!\!\mod #1)}
\newcommand{\half}{\tfrac12}
\newcommand{\Tr}[1]{{\rm Tr}[#1]}
\font\small=cmr8 scaled \magstep0

\font\grande=cmr10 scaled \magstep4
\font\medio=cmr10 scaled \magstep2
\outer\def\beginsection#1\par{\medbreak\bigskip
      \message{#1}\leftline{\bf#1}\nobreak\medskip
\vskip-\parskip
      \noindent}

\newcommand{\bi}{\begin{itemize}}
\newcommand{\ei}{\end{itemize}}
\newcommand{\eq}{\begin{equation}}
\newcommand{\eqx}{\end{equation}}
\newcommand{\eqn}{\begin{eqnarray}}
\newcommand{\eqnx}{\end{eqnarray}}
\newcommand{\ad}{{a^{\dagger}}}
\newcommand{\fd}{f^{\dagger}}

\newcommand{\Qd}{{Q^{\dagger}}}

\newtheorem{thm}{Theorem}

\newcommand{\ra}{\rangle}

\begin{document}

\titlepage

\begin{flushright}
\vspace{15mm}
CERN-PH-TH/2006-046\\
TPJU-6/2006\\
\end{flushright}
\vspace{15mm}
\begin{center}

\grande{Supersymmetry and Combinatorics$^*$}

\vspace{15mm}

 \large{E. Onofri}

   \vspace{5mm}

   {\sl Dipartimento di Fisica, Universit\`a di Parma}

 {\sl and}

{\sl I.N.F.N., Gruppo Collegato di Parma, 43100 Parma, Italy}

\vspace{5mm}

\large{G. Veneziano}

\vspace{5mm}

 {\sl Theory Division, CERN, CH-1211 Geneva 23, Switzerland }

{\sl and}

{\sl Coll\`ege de France, 11 place M. Berthelot, 75005 Paris, France}
\vspace{10mm}

   \large{J. Wosiek}

   \vspace{5mm}

   {\sl M. Smoluchowski Institute of Physics, Jagellonian University}

{\sl Reymonta 4, 30-059 Cracow, Poland}

\end{center}

\centerline{\medio  Abstract}
\vskip 5mm
\noindent
We show how a recently proposed supersymmetric quantum mechanics model
leads to non-trivial results/conjectures on the combinatorics of
binary necklaces and linear-feedback shift-registers. 
Pauli's exclusion principle plays a crucial role: by projecting out certain
states/necklaces, it allows to represent
the supersymmetry algebra in the resulting  subspace.
Some of our results can be rephrased in terms of generalizations of
the well-known Witten index. 
 \vspace{5mm}
 
*) Revised version, December 2006

\vfill
\newpage

\section{Introduction}
In a recent series of papers \cite{VW1}--\cite{VW2} two of us have
introduced a supersymmetric quantum mechanical matrix model and
studied some of its intriguing properties.  The model is defined as
the $N \rightarrow \infty$ limit of a quantum mechanical system whose
degrees of freedom are bosonic and fermionic $N \times N$ creation and
destruction operator matrices. The model's supersymmetry charges and
Hamiltonian are explicitly given by: 
\eq 
Q= \Tr{f \ad(1+g\ad)},
\;\;\; \Qd= \Tr{\fd (1+g a) a}, \;\;\; Q^2 = \Qd^2 =0 \, , 
\eqx
\eq 
H=\{Q^{\dagger},Q\} = H_B+H_F \, , \label{h1} 
\eqx 
\eq 
H_B= \Tr{\ad a + g(\ad^2 a + \ad a^2) + g^2 \ad^2 a^2} \, , \label{h2}
\eqx 
\eqn
H_F&=& {\rm Tr} [\fd f + g ( \fd f (\ad+a) + \fd (\ad+a) f) \nonumber \\
& + & g^2 ( \fd a f \ad + \fd a \ad f + \fd f \ad a + \fd \ad f a)] \,
,\label{h3} \eqnx 
where bosonic and fermionic destruction and creation operators satisfy
\eq 
[a_{ij},\ad_{kl}]=\delta_{il}\delta_{jk} \,~ ; \,~ \{f_{ij}
\fd_{kl}\}=\delta_{il}\delta_{jk} \, ; \,\,\, i,j,k,l =1,\dots N \, ,
\label{com}
\eqx 
all other (anti)commutators being zero. While taking the large-$N$
limit, one keeps, as usual \cite{tH}, the 't Hooft coupling, $\lambda
\equiv g^2 N$, fixed. Note that the Hamiltonian (\ref{h3}) conserves
(commutes with) the fermionic number $F=\Tr{\fd f}$. Hence the
system can be studied separately for each eigenvalue of $F$. By
contrast, $H$ does not commute with the bosonic number operator
$B=\Tr{\ad a}$ except in the trivial $g \rightarrow 0$ limit.

The model exhibits  a number of interesting properties:

\begin{enumerate}
\item It is exactly soluble in the $F=0,1$ sectors, i.e.  the
  complete energy spectrum and the eigenstates are available in
  analytic form, in particular 
  it exhibits a discontinuous phase transition at $\lambda=
  \lambda_{\rm c} = 1$. At this point the otherwise discrete spectrum
  loses its energy gap and becomes continuous.
\item An exact weak-strong duality holds in the $F=0,1$ sectors
  relating spectra at $\lambda$ and $1/\lambda$.
\item It exhibits unbroken supersymmetry, i.e. its $E \ne 0$
  eigenstates consist of degenerate boson-fermion doublets.
\item In the weak coupling phase, $\lambda < 1$, there is only one
  (unpaired) zero-energy state (also referred to as a SUSY vacuum). It
  lies in the $F=0$ sector and is nothing else but the empty Fock
  state $|0\rangle$ while for $\lambda > 1$ there are {\em two}
  zero-energy states in each bosonic (even $F$) sector of the
  model. For $F=0$ the Fock vacuum continues to be a zero energy
  eigenstate, but it is joined by another, non-trivial, analytically
  known ground state. For each higher even $F$, the two non-trivial
  ``vacua'' appear suddenly at $\lambda > 1$. Some understanding of
  these unexpected states was obtained by considering the $\lambda
  \rightarrow \infty$ limit of the model \cite{VW3}.  In that same
  limit, the model can be connected to two interesting one-dimensional
  statistical mechanics quantum systems \cite{VW3}.
\end{enumerate}

  In the appropriate large-$N$ limit defined above, the Hilbert space
  of the model can be restricted to the one corresponding to the
  action of single-traces of products of creation operators acting on
  the Fock vacuum. As such the vectors of the large-$N$ Hilbert space
  can be put in one-to-one correspondence with binary necklaces, with
  the two beads representing bosonic and fermionic matrices. However,
  Fermi statistics provides a well-defined ``Pauli razor'', which
  projects out a subset of all binary necklaces. As we shall see, only
  after this projection is performed, the resulting space does allow
  for a faithful representation of supersymmetry\footnote{After this
  work was completed we learned from M.  Bianchi that some of the
  results presented here had already been derived (or guessed) by
  other methods in Refs. \cite{Bianchi}. We wish to thank M. Bianchi
  for the information and for instructive discussions about that
  work.}.

  The purpose of this paper is to illustrate how supersymmetry in our
  physical model gives non-trivial results on the combinatorics of
  binary necklaces and how, vice versa, known combinatorics results on
  the latter allow to determine the way supersymmetry is realized.  In
  particular, combinatorics will allow us to understand where the null
  eigenstates lie and to compute the value of the Witten index
  \cite{WI} --and generalizations thereof-- in different regions of
  $\lambda$.

  The rest of the paper is organized as follows: in Sec.~2~ we explain
  how single-trace states are connected to necklaces and describe how
  they can be enumerated taking into account Fermi statistics; the
  concept of Pauli {\sl allowed\/} or {\sl forbidden necklaces\/} is
  introduced together with some examples. We also introduce there the
  connection with ``linear feedback shift registers'' which helps in
  finding the correct answer.  In Sec.~3~ we provide a generalization
  of Polya's formula, by giving the number of forbidden necklaces with
  a prescribed number of bosonic and fermionic beads. In Sec.~4~ we
  show how supersymmetry suggests combinatorial identities, which can
  also be proven by classical arguments, but are otherwise difficult
  to envisage. The Appendix provides a technical proof of a corollary
  of the main theorem.

\section{Fock states, necklaces and linear feedback registers}

The Hilbert space of our model (\ref{h1}) is spanned by states created
 by single trace operators, e.g.
\begin{equation*}
  \Tr{a^\dagger\ad  f^\dagger ...  \ad a^\dagger \fd f^\dagger} |0\ra \, .
\end{equation*}
These are also eigenstates of the above-mentioned fermion and boson
number operators $F$ and $B$.  Such states too can be labeled by
binary numbers, e.g.  
\eq 
(0 0 1 ... 0 0 1 1 ), \label{binary} 
\eqx
with 0 (1) corresponding to bosonic (fermionic) creation
operators. Because of cyclic property of a trace, all $n$ binary
sequences, related by cyclic shifts, describe the same state with $n$
quanta, and consequently should be identified. Therefore our states
correspond to what mathematicians define as necklaces - periodic
chains made of different beads. In our model only two kinds of beads
occur, hence only {\em binary} necklaces will be encountered
here. From now on, if not specified otherwise, a term necklace will
mean a binary necklace.

Since we are dealing with fermions, some of the above necklaces will
not be allowed by the Pauli exclusion principle which will turn out to
be crucial for supersymmetry as already mentioned in the
introduction. We therefore define the {\em allowed} and {\em
forbidden} necklaces as those which are allowed and forbidden by the
Pauli principle. Hence the set of all necklaces is the union of
allowed and forbidden ones.

\subsection{Counting states/necklaces}

We begin by recalling the classical results on counting all binary
necklaces.  The total number $N(n)$ of necklaces with $n$ beads is
given by MacMahon's formula
\begin{equation}
  \label{McMAhon}
  N(n)  = \frac1n \sum_{d|n}\, \varphi(d)\,2^{n/d}\,,
\end{equation}
where $d|n$ means that $d$ divides $n$ and $\varphi(d)$ is Euler's
``totient'' function, counting the numbers in $1,2, ... , d-1$
relatively prime to $d$.  The slightly "more differential" number
number $N(B,F)$ of necklaces with $B$ and $F$ separate beads ($B$
beads of type ``0'' and $F$ of type ``1'') is given by Polya's formula
\cite{lint92,stanley99}:

\begin{equation}
  \label{Polya}
  N(B,F)=\frac1{B+F}\,\sum_{d|B,F} \,\varphi(d)\,{B/d+F/d\choose
F/d}\, ,
\end{equation}
where $d|B,F$ means that $d$ is a divisor of both $B$ {\sl and\/} $F$.
Of course the numbers given in (\ref{Polya}) sum up to those in
(\ref{McMAhon}).

\subsection{Allowed vs. forbidden necklaces}
Let us now look in more detail how the antisymmetry excludes some of
the planar states/necklaces.  For instance the sequence $0101$
corresponding to the operator
\begin{equation*}
 \Tr{a^\dagger f^\dagger a^\dagger f^\dagger}
\end{equation*}
vanishes identically, since by anticommutation of $f$ one has
\begin{equation*}
\Tr{a^\dagger f^\dagger a^\dagger f^\dagger}=- 
\Tr{f^\dagger a^\dagger f^\dagger a^\dagger}
  = -\Tr{a^\dagger f^\dagger a^\dagger f^\dagger} = 0
\end{equation*}
On the other hand the sequence $aaff$ survives, since
\begin{equation*}
  \Tr{a^\dagger a^\dagger f^\dagger f^\dagger}=- \Tr{ f^\dagger a^\dagger a^\dagger f^\dagger}
  = +\Tr{f^\dagger f^\dagger a^\dagger a^\dagger} = + \Tr{a^\dagger a^\dagger f^\dagger f^\dagger}
\end{equation*}

Our problem is to find in a systematic way which necklaces and how
many of them survive the Pauli principle. The distinction into allowed
and forbidden necklaces crucially depends whether a necklace has even
or odd number of fermionic quanta.  Therefore, from now on, we reserve
the term {\em fermionic necklace} to one with an odd number of
fermions (1's in its binary representation), while a {\em bosonic
  necklace} will denote a necklace with even number of fermionic
quanta $F$.  It follows that fermionic necklaces are always allowed,
since a cyclic shift consists of even number of fermionic
anticommutations, while some of the bosonic necklaces may be
forbidden. To see this consider two longer necklaces 
\eq
(011011),\;\;\;\; {\rm and}\;\;\;\; (01010101) 
\eqx 
both of them are bosonic, however only the first one is allowed. Both
states have an additional symmetry ($Z_2$ and $Z_4$, respectively), 
but the number of fermionic transpositions needed to shift them into
themselves is different. A little thought allows now to identify the
necessary and sufficient condition for a necklace to be forbidden:

\vspace{1mm} {\bf A necklace with $Z_{k}$ symmetry, $k$ even, and
  $\frac{F}{k}$ odd, is forbidden and vice versa.}  \vspace{1mm}
\noindent Because of the cyclic invariance of a trace and of Fermi
statistics, we find that these states are equal to their opposite and
hence vanish.

The above condition splits the space of all necklaces in the way
sketched in Fig.~1 where the necklaces are divided into four groups
according to whether they contain even or odd number of fermionic and
bosonic beads.  The exclusion principle is only effective in the
even-even group where some necklaces are Pauli-forbidden.

Supersymmetry manifests itself in terms of the existence of doublets
of energy eigenstates (with non-vanishing eigenvalue) consisting of a
boson (a bosonic necklace) and a fermion (a fermionic necklace).
Precisely the removal of the forbidden necklaces from the even-even
sector should give back the balance between even-even and odd-odd
sectors required by supersymmetry.
\begin{figure}[ht]
\begin{center}
\includegraphics[width=1.0\textwidth,clip=true ]{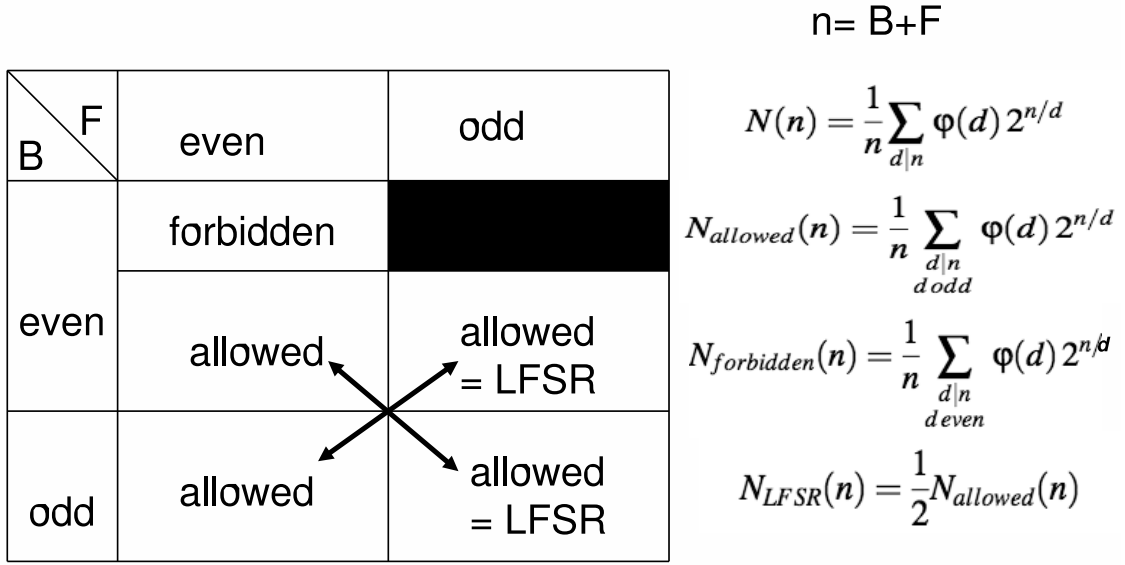}
\end{center}
\caption{A map of the space of all necklaces for even and odd $B$ and $F$
  showing where  Pauli-forbidden necklaces lie. The
  double arrows indicate the way supersymmetry connects allowed
  necklaces of opposite statistics at weak coupling. The connection
  with linear feedback shift registers (LFSR) is explained in
  subsection 2.3.}
\end{figure}

\subsection{Allowed necklaces and linear feed-back registers}

It turns out that fermionic necklaces are closely related to linear
feedback shift registers (LFSR) - yet another class of objects well
known in combinatorics \cite{LFSR}. In order to avoid lengthy
definitions we illustrate hereafter the concept of a LFSR, of length
$n$, and its relation to odd necklaces of length $n$ (giving $n=4$ as
an example):

i) Definition of a LFSR

\begin {itemize}
\item Take an arbitrary binary number with $n-1$ (here 3) digits: 
\eq
  (000, 001, 010, 011, \dots , 111); \nonumber 
\eqx
\item Start adding digits to its right by the following (linear
  feedback) rule: add a 0 if the sum of the three digits is odd and a
  1 if the sum is even. This gives: 
\eq 
(0001, 0010, 0100, 0111, \dots
  , 1110); \nonumber 
\eqx 
By construction, the sum of the {\bf 4}
  digits is always odd.
\item Repeat the procedure by applying the rule to the new last three
  figures. The result is 
\eq (00010, 00100, 01000, 01110, \dots , 11101) \, . \nonumber 
\eqx 
Clearly the 5th figure coincides with the first. If we keep going, we
get a series that is periodic with period {\bf 4}.
\end {itemize}

ii) Claim of equivalence

The claim is that the distinct LFSR thus obtained are in one-to-one
correspondence with fermionic necklaces of length $n$.

Proof: we have already argued that elementary cells of length $n$ have
an odd sum.  Also, if two cells of length $n$ are related by a cyclic
transformation, they lead to the same infinite periodic
structure. Thus every inequivalent, odd necklace of length $n$ gives a
distinct infinite sequence of period $n$ and vice versa.
 
 This proof is illustrated by the following three infinite sequences:
 \begin{eqnarray}
 &&010001000100010001 \dots \nonumber \\
  &&100010001000100010  \dots \nonumber \\
   &&110111011101110111 \dots 
   \label{seq}
 \end{eqnarray}
 generated by our rule out of three different initial ``data''. The
 first two infinite sequences are considered to be equivalent: they
 correspond to the same 4-digit periodic structure (up to a cyclic
 permutation) repeating itself indefinitely, and corresponds to the
 odd necklace of fig. 2 (left side), while the third sequence gives the only
 other inequivalent odd necklace of length four, also shown in fig. 2 (right side).

\begin{figure}[ht]
  \includegraphics[width=.5\textwidth,clip=true]{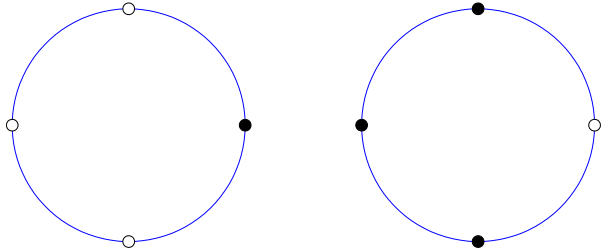}
 \caption{ The two inequivalent necklaces corresponding to the infinite sequences  (\ref{seq}).}
 \label{fig:?}
 \end{figure}

The number of linear feed-back registers of length $n$ is
catalogued as A000016(n) in Sloane's library (we
shall often refer to this remarkable tool \cite{sloane}, which was
very helpful at a certain stage of our work). In conclusion

\begin{equation}\label{eq:Nferm}
  N_{\rm{fermionic}}(n)  = N_{\rm{LFSR}}(n) = A000016(n)  =
  \frac1{2n} \sum_{\begin{subarray}{c}
      d|n\\
      d\, {\rm odd}
    \end{subarray}}
  \,\varphi(d)\,2^{n/d} \, .
\end{equation}

\subsection{ Separate, global counting of allowed and forbidden necklaces}

Supersymmetry requires that the numbers of allowed bosonic and
fermionic necklaces are the same for given $n$. However this general
condition has different consequences for even and odd $n$.  If $n$ is
odd all necklaces are allowed (c.f. Fig.1) and consequently the number
of bosonic necklaces coincides with the number of fermionic ones. In
this case the restriction ``$ d $ odd'' in the last equation is
superfluous and we get the total count of necklaces given by
MacMahon's formula (\ref{McMAhon}). On the contrary, when $n$ {\sl is
even\/} we still have the value of $N_{\rm{fermionic}}$ given by
Eq.~\eqref{eq:Nferm}, but now supersymmetry requires this to match the
number of {\em allowed} bosonic necklaces in the even-even
sector. This yields the general (i.e. valid for all $n$) rule

\begin{equation}
  \label{allowed}
  N_{\rm allowed}(n) =   \frac1{n} \sum_{\begin{subarray}{c}
      d|n\\
      d\, {\rm odd}
    \end{subarray}}
  \,\varphi(d)\,2^{n/d} \, ,
\end{equation}
implying a total number of forbidden necklaces
\begin{equation}
  \label{forbidden}
  N_{\rm {forbidden}}(n) =   \frac1{n} \sum_{\begin{subarray}{c}
      d|n\\
      d\, {\rm even}
    \end{subarray}}
  \,\varphi(d)\,2^{n/d} \, ,
\end{equation}
for any $n$.  For small even n ($2\le n \le 32$), the formula gives 
$N_{\rm{forbidden}}(n) = 1, 2, 2, 4, 4, 8, 10,\\ 20, 30, 56, 94, 180, 316,
  596, 1096 $.

This result follows by supersymmetry; it will be also derived by
traditional combinatorial arguments later on (see Appendix).  It is
perhaps amusing that the obvious algebraic fact that odd $n$ has only
odd divisors while even $n$ admits both (i.e. even and odd) divisors,
directly corresponds to the existence of the necklaces allowed and
forbidden by the Pauli principle!

\section{ Generalization of Polya's formula for forbidden necklaces}

We would like to find the counterpart of Polya's formula
eq.~(\ref{Polya}) which holds for the bosonic/fermionic necklaces with
fixed numbers of separate beads . Equivalence between classical
necklaces and susy necklaces when $F$ is odd tells us that, in this
case, we simply have:
\begin{eqnarray}
  \label{PANs}
  N_{\rm allowed}(B,F)&=&\frac1{B+F}\,\sum_{\begin{subarray}{c}
      d|B,F\\
    \end{subarray}} \,\varphi(d)\,{B/d+F/d\choose F/d} \nonumber \\
    & =& \frac1{B+F}\,\sum_{\begin{subarray}{c}
      d|B,F\\
      d\, {\rm odd}
    \end{subarray}} \,\varphi(d)\,{B/d+F/d\choose F/d} \, , \, F~ {\rm
    odd}\, .
     \end{eqnarray}
     By an obvious symmetry, the same formula holds if $B$ is odd and
     $F$ is even.  The only tricky case, again, is the one where both
     $B$ and $F$ are even: here we want to distinguish allowed from
     forbidden necklaces and count them separately for given values of
     $B$ and $F$.

     It turns out to be easier to find first the general formula for
     the number of forbidden necklaces, which, when combined with
     Polya's eq.~(\ref{Polya}), will produce as a corollary also the
     number of allowed necklaces.  Our claim is as follows:
\begin{thm}
  Let $r$ be the unique positive integer (if it exists) for which
  $f=F/2^r$ is odd and $b=B/2^r$ is an integer.  The number of Pauli
  Forbidden Necklaces is given by
\begin{equation}
  \label{PFNsBF}
  N_{\rm{forbidden}}(B,F)=
  N(B/2^r,F/2^r)  =  \frac1{b+f}\,\sum_{d|b,f}
  \,\varphi(d)\,{b/d+f/d\choose f/d}\,.
    \end{equation}
    If such an $r$ does not exist then $N_{\rm{forbidden}}(B,F)=0$.
\begin{proof}
  Pauli principle is active in deleting necklaces which are
  $Z_p$-symmetric with $p$ even and $F/p$ odd; then it is clear that,
  by writing $p=2^r q$ with $q$ odd, $F/2^r =f$ must be odd and $B/2^r
  = b$ must be an integer. If we now consider any sequence of length
  $b+f$ (a cell repeated $2^r$ times along the whole necklace), we see
  that such a cell is itself an arbitrary necklace with $b$ bosons,
  $f$ fermions and symmetry $Z_q$ with $q$ any odd number. Since $f$
  is odd, such a $Z_q$ symmetry covers all possible cases, and
  therefore the number of inequivalent cells is indeed given by
  Polya's formula; notice that a different cyclic permutation of the
  elementary cell gives the same necklace, because a cyclic
  permutation of the cell is equivalent to a cyclic permutation of the
  whole necklace.
\end{proof}
\end{thm}

As an example, consider the string $\Tr{\fd \ad \ad \fd \ad
  \ad \fd \ad \ad \fd \ad \ad}$ corresponding to the necklace of Fig.3.

\begin{figure}[ht]
 \includegraphics[width=.65\textwidth,clip=true]{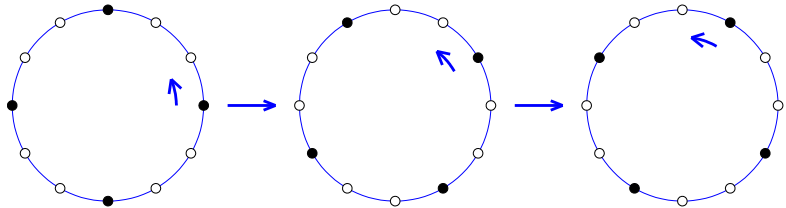}

 \caption{{\small Counterclockwise rotating the necklace until it
      matches the initial configuration (from left to right): $(f~ a~ a~ f~ a~ a~ f~ a~ a~ f~a~a), (a~f~ a~a~ f~ a~ a~ f~ a~ a~
     f~ a), (a~a~ f~ a~ a~ f~a~ a~ f~ a~ a~ f)$}}
 \label{fig:2}\end{figure}
The elementary cell in this case is $(faa)$ and the number of
forbidden necklaces coincides with $N(2,1)=1$, which corresponds
to the necklaces depicted in Fig.3. In fact, rotating the necklace
until it matches the initial configuration, one has an odd number of
fermionic commutations yielding the minus sign which kills the
necklace.

Finally, by taking the difference between eq.~(\ref{Polya}) and
(\ref{PFNsBF}), we  conclude that:
\begin{eqnarray}
  \label{PANsBF}
  N_{\rm allowed}(B,F) = N (B,F)-N(B/2^r,F/2^r)\, ,
 \end{eqnarray}
 where $r$ is as defined in Theorem 1.
 The values of $N_{\rm allowed}(B,F)$ for $F+B \le 26$ are
 reported\footnote{This table and the following one were produced by a
   {\sl sieve} method for $B+F\le 26$, independently of Theorem 1, and
   were used to check the general formulae.}  in the form of a $(B,F)$
 array in table~\ref{tab:BF}.  At this point, if our
 supersymmetry-based argument is correct, summing over the number of
 allowed necklaces given by eq.~(\ref{PANsBF}) for fixed $n$ should
 reproduce exactly the same number as twice the sum over the fermionic
 supersymmetric partners, i.e. just eq.~(\ref{allowed}).  We have
 verified numerically that this is the case up to $n =3000$ and later
 found a direct mathematical proof reported in the appendix.  The
 existence of such a proof confirms the solidity of the
 supersymmetry-based arguments, as well as their considerable
 heuristic value.

\section{Witten-like indices}

The formula for $N_{\rm allowed}$ verifies a number of checks coming
from the properties of the supersymmetric model at weak and strong
coupling. In the first, weak-coupling regime, which is fully under
control, supersymmetry tells us that the allowed necklaces with a given
$n=B+F$ should organize themselves in supersymmetry doublets, each of
which consists of a necklace with some $B$ and $F$ and one with $B' =
B\pm1$ and $F' = F\mp 1$.  Since the number of such pairs is always
non-negative, we obtain the following inequalities for graded partial
sums (a kind of generalization of Witten's index \cite{WI}):
\begin{equation}
  W(n;m) \equiv \sum_{\begin{subarray}{c}
        B+F=n\\
        0\le F \le m
      \end{subarray}}(-1)^{F-m} \,N_{\rm allowed}(B,F) \ge  0 ~~, ~~ W(n;n) =0 \, ,
\end{equation}
where the last equality corresponds to that between even and odd
allowed necklaces with a given $n$.  The above consequences of
supersymmetry have been explicitly checked up to $n \sim 5000$),
while, so far, we have not been able to construct a direct proof of
them by more standard techniques.

The strong ('t Hooft) coupling limit of the model of \cite{VW1} can be
shown \cite{VW3} to imply instead that allowed necklaces must also
organize in supersymmetry doublets whose partners have the same value
of $B+2F$ (and again differ by one, positive or negative, unit of
$F$). A look at Table 1 shows that, along diagonals at fixed $B+2F$,
the balance between even and odd allowed necklaces is not always
satisfied. This implies that, along those diagonals, there must be, at
large coupling, (unpaired) $E=0$ states.

The large-coupling limit unfortunately is not fully under control
yet. Therefore, in this case, the connection between eigenstates and
allowed necklaces can be used in either direction to infer properties
of one in terms of known properties of the other. For instance, some
evidence has been accumulated on where zero-energy states lie in the
$B,F$ plane.  On the basis of this evidence we can conjecture new
checks on our formulae for $N_{\rm allowed}$ by the following property of
a second Witten-like index:
 \begin{equation}
   \label{PSTRSC}
  \tilde{W}(n;m) \equiv \sum_{\begin{subarray}{c}
        B+2F=n\\
        F \le m/2
      \end{subarray}}(-1)^{F-[m/2]} \,\left( N_{\rm allowed}(B,F)-
      \frac{\delta_{F,B+1} + \delta_{F,B-1}}{2} (1 +(-1)^F) \right)
    \ge 0 \, ,
\end{equation}
(with $m \le n$) and, in particular,
\begin{equation}
  \label{STRSC}
  \tilde{W}(n ; n)=0 \Rightarrow \sum_F(-1)^FN_{\rm allowed}(n-2F,F) = \delta_{n\equiv 1\Mod6}+
  \delta_{n\equiv -1\Mod6}\,.
\end{equation}
Our  formulae passed the test of these (in)equalities for $n\le 5000$.
 
 Actually, the validity of eq.  (\ref{STRSC}) for all values
of $n$ follows from an explicit expression for the generating function of $N_{\rm
  allowed}(B,F)$ recently obtained by D. Zagier:
\begin{equation*}
 \Phi_{\rm allowed}(x,y;n)\;\equiv \;\sum_{F=0}^n N_{\rm allowed}(n-F,F) \,x^{n-F}y^F
 =\frac1n\,\sum_{d|n}\,\varphi(d)\,\left(x^d-(-y)^d\right)^{n/d}\,,
\end{equation*}
 after setting $y= -x^2$ and summing over $n$ \footnote{We are very grateful to Professor Zagier  for informing us of this  result, and for giving us
permission to report it here.}.

When $B+2F$ is small, the zero-energy eigenstates causing the
imbalance can be uniquely identified in table~\ref{tab:BF}, while, for
the moment, their identification can only be guessed at (and verified
later) for $B+2F$ large. This is how we arrived at the conjecture
\cite{VW3} that, at strong coupling, there is one and only one
zero-energy eigenstate for each even value of $F$ and $B = F \pm1$, a
conjecture leading precisely to eqs.~(\ref{PSTRSC}) and (\ref{STRSC}).

Finally, it is amusing to notice that the total number of
strong-coupling eigenstates at these special locations (forming a kind
of magic staircase in table~\ref{tab:BF}) is given by the sequence
\begin{equation*}
  1,1,2,5,14,42,132,429,1430,4862,16796,58786,\ldots
\end{equation*}
which is easily recognized as being that of Catalan's numbers:
\begin{equation*}
  N_{{\rm Catalan}} = \frac1{n+1}{2n\choose n} \, .
\end{equation*}
Catalan's numbers are ubiquitous, 66 appearances of them being listed
in Stanley's treatise \cite{stanley99}. It is easy to convince oneself
that to every necklace with $|B-F|=1$ one can associate an infinite
sequence of {\sl ups} and {\sl downs} describing a mountain profile,
the number of which is precisely given by Catalan numbers
\cite{conway96}. These entries belong to the subset with $N_{\rm
forbidden}=0$, since either $B$ or $F$ is odd. Other diagonals can be
identified with known sequences; for instance, $N_{\rm
allowed}(F\pm2,F)$ is identical to the number of plane trees with
odd/even number of leaves (A071684, A071688 \cite{sloane}).

To summarize our main results:
\begin{itemize}
\item We have been able to divide all binary necklaces  in two disjoint classes, which we termed (Pauli)-allowed and (Pauli)-forbidden .
\item At the most ``inclusive'' level, the number of binary necklaces with a total
  number $n$ of beads, as given by MacMahon's formula (\ref{McMAhon}),
  is split into allowed and forbidden necklaces by restricting the divisor $d$ in
  (\ref{McMAhon}) to odd and even values, respectively.
\item At a more ``differential'' level, the number of necklaces with $B$
  bosonic and $F$ fermionic beads is rewritten in terms of allowed
  necklaces with different values of $B$ and $F$ via
  eq.~(\ref{PANsBF}), which can also be rewritten as:
\begin{eqnarray}
  \label{BNLsBF}
  N(B,F) &=& N_{\rm allowed}(B,F)~ + ~ N(B/2^r,F/2^r) \, , \\ \nonumber
  N(B/2^r,F/2^r)  &=& N_{\rm allowed}(B/2^r,F/2^r) \, ,
\end{eqnarray}
where $r$ is as defined in Theorem 1. We have verified numerically (and
then proved directly, see appendix) that the appropriate sum performed
on (\ref{BNLsBF}) reproduces the above-mentioned relation at fixed $n
= B+F$.
\item Supersymmetry implies several non-trivial constraints on $N_{\rm
    allowed}(B,F)$ and thus, through (\ref{BNLsBF}), also on $N(B,F)$.
     Examples have been given in Section 4, but we stress
  that, by suitably extending the supersymmetric model under
  consideration, it is quite conceivable that many more constraints
  will emerge, not only for binary necklaces, but also for their generalization to
  more than two kinds of beads.
\end{itemize}
This new  game (that we may dub ``super-combinatorics'')  should reserve further surprises both for physicists and for mathematicians.

\section*{Acknowledgements}
GV would like to acknowledge interesting discussions with Professors
Jean-Christophe Yoccoz and Don Zagier. JW thanks A.  Kotanski for
instructive discussions. EO warmly thanks G. Cicuta for interesting
discussions. This work is partially supported by the grant of the
Polish Ministry of Education and Science P03B 024 27 (2004--2007).

\section*{Appendix: Proof of consistency between eqs. ~(\ref{forbidden})
  and (\ref{PFNsBF}).}

\begin{thm}
$$
\sum_{F=0}^n N_{\rm forbidden}(n-F,F)= \frac1n\sum_{\substack{d|n\\ d\, {\rm even}}}\phi(d)2^{n/d} \, .
$$
\begin{proof}
Let $n=2^rq$, with $q$ odd. We have
\begin{eqnarray*}
  &&\frac1{2^rq}\sum_{\substack{d|2^rq\\d\, {\rm even}}}\phi(d)2^{2^rq/d}
= \frac1{2^rq}\sum_{m=1}^r\sum_{d|q}\phi(2^md)2^{2^{r-m}q/d}\\
&=& \frac1{2^rq}\sum_{m=1}^r2^{m-1}\sum_{d|q}\phi(d)2^{ 2^{-m}n/d}
= \half\sum_{m=1}^r\frac{2^m}{n}\sum_{d|q}\phi(d)2^{2^{-m}n/d}
= \sum_{m=1}^r N_{\rm LFSR}(2^{-m}n)
  \end{eqnarray*}
(we use  $\phi(2^m)=2^{m-1}$
and in the second step we put $d=d_1d, d_1|2^r, d|q$).

On the other hand the only non-vanishing contributions to $
\sum_{F=0}^nN_{\rm forbidden}(n-F,F)$ come from $F=2^mq'$, where $m=1,2,...,r$ and
$q'=1,3,5,...,q$,  so that we have:
\begin{eqnarray*}
&& \sum_{F=0}^nN_{\rm allowed}(n-F,F) = \sum_{F=0}^nN_{\rm forbidden}(2^rq-F,F)
= \sum_{m=1}^r\sum_{\substack{q'=1\\q'\,{\rm odd}}}^qN_{\rm forbidden}(2^rq-2^mq',2^mq')\\
&=& \sum_{m=1}^r\sum_{\substack{q'=1\\q'\,{\rm odd}}}^qN_{\rm BNL}(2^{r-m}q-q',q')
= \sum_{m=1}^r N_{\rm LFSR}(2^{-m}n) \, .
\end{eqnarray*}

\end{proof}

\end{thm}

\newpage

\pagestyle{empty}

\begin{landscape}
{\small
\begin{table}[b]
   \begin{center}
     \begin{tabular}{l||r|r|r|r|r|r|r|r|r|r|r|r|r|r|r|r|r|r|r|r|r|}
      & \multicolumn{21}{l}{\mbox{$F\rightarrow$}}\\\hline
      $B\downarrow$ & 0& 1&2 &3 &4 &5 &6 &7 &8 &9 &10 &11 &12 &13 &14 &15 &16 &17&18&19&20 \\ \hline\hline
0&1&1&0&1&0&1&0&1&0&1&0&1&0&1&0&1&0&1&0&1&0\\\hline
1&1&1&  1&  1&  1&  1&  1&  1&  1&  1&  1&  1&  1&  1&  1&  1&  1&  1&  1&  1&  1\\\hline
2&1&1&  1&  2&  3&  3&  3&  4&  5&  5&  5&  6&  7&  7&  7&  8&  9&  9&  9& 10& 11\\\hline
3&1&1&  2&  4&  5&  7& 10& 12& 15& 19& 22& 26& 31& 35& 40& 46& 51& 57& 64& 70& 77 \\\hline
4&1&1&  2&  5&  9& 14& 20& 30& 43& 55& 70& 91&115&140&168&204&245&285&330&385&445 \\\hline
5&1&1&  3&  7& 14& 26& 42& 66& 99&143&201&273&364&476&612&776&969& 1197& 1463& 1771& 2126   \\\hline
6&1&1&  3& 10& 22& 42& 76&132&217&335&497&728& 1038& 1428& 1932& 2586& 3399& 4389& 5601& 7084& 8866    \\\hline
7&1&  1&  4& 12& 30& 66&132&246&429&715& 1144& 1768& 2652& 3876& 5538& 7752&10659&14421&19228&25300 &    \\\hline
8&1&  1&  4& 15& 42& 99&212&429&809& 1430& 2424& 3978& 6308& 9690&14520&21318&30667&43263&60060&  &    \\\hline
9&1&  1&  5& 19& 55&143&335&715& 1430& 2704& 4862& 8398&14000&22610&35530&54484&81719&  120175  &  & &\\\hline
10&1&  1&  5& 22& 73&201&497& 1144& 2438& 4862& 9226&16796&29414&49742&81686&  130752&  204347 &  &  &&   \\\hline
11&1&  1&  6& 26& 91&273&728& 1768& 3978& 8398&16796&32066&58786&  104006&  178296&  297160&    &  &  & &  \\\hline
12&1&  1&  6& 31&115&364& 1028& 2652& 6310&14000&29372&58786&  112716&  208012&  371384&  &  &  &  &    &\\\hline
13&1&  1&  7& 35&140&476& 1428& 3876& 9690&22610&49742&  104006&  208012&  400024&  &  &    &  &  &   &\\\hline
14&1&  1&  7& 40&172&612& 1932& 5538&14550&35530&81686&  178296&  371516&  &  &  &  &  &    &    &\\\hline
15&1&  1&  8& 46&204&776& 2586& 7752&21318&54484&  130752&  297160&  &  &  &  &  &  & & &     \\\hline
16&1&  1&  8& 51&244&969& 3384&10659&30666&81719&  204248&  &  &  &  &  &  &  &  &  &  \\\hline
17&1&  1&  9& 57&285& 1197& 4389&14421&43263&  120175&  &  &  &  &  &  &  &  &  &  &   \\\hline
18&1&  1&  9& 64&335& 1463& 5601&19228&60115&  &  &  &  &  &  &  &  &  &  &  &    \\\hline
19&1&  1& 10& 70&385& 1771& 7084&25300&  &  &  &  &  &  &  &  &  &  &  &  &   \\\hline
20&1&  1& 10& 77&445& 2126& 8844&  &  &  &  &  &  &  &  &  &  &  &  &  &  \\\hline
21&1&  1& 11& 85&506& 2530&  &  &  &  &  &  &  &  &  &  &  &  &  &  &    \\\hline
22&1&  1& 11& 92&578&  &  &  &  &  &  &  &  &    &  &  &  &  &  &  &    \\\hline
23&1&  1& 12&100&  &  &  &  &  &  &  &  &  &  &  &  &  &  &  &  &   \\\hline
24&1&  1& 12&  &  &  &  &  &  &  &  &  &   &  &  &  &  &  &  &  &  \\\hline
25&1&  1&  &  &  &  &  &  &  &  &  &  &    &  &  &  &  &  &  &  &   \\\hline
26&1&  &  &  &  &  &  &  &  &  &  &  &   &  &  &  &  &  &  &  &  \\\hline
      \end{tabular}\\[.5em]
    \end{center}
 \caption{$N_{\rm allowed}(B,F)$ as generated with the sieve method.}\label{tab:BF}
\end{table}

\begin{table}[h]
  \begin{center}
    \begin{tabular}{l||r|r|r|r|r|r|r|r|r|r|r|r|r|r|r|r|r|r|r|}
&\multicolumn{19}{l}{$F\rightarrow$}\\\hline
$B\downarrow$&0&2&4&6&8&10&12&14&16&18&20&22&24&26&28&30&32&34&36\\\hline\hline
0& 0 & 1 & 1 & 1 & 1 & 1 & 1 & 1 & 1 & 1 & 1 & 1 & 1 & 1 & 1 & 1 & 1 & 1 & 1  \\\hline
2& 0 & 1 & 0 & 1 & 0 & 1 & 0 & 1 & 0 & 1 & 0 & 1 & 0 & 1 & 0 & 1 & 0 & 1 & 0  \\\hline
4& 0 & 1 & 1 & 2 & 0 & 3 & 1 & 4 & 0 & 5 & 1 & 6 & 0 & 7 & 1 & 8 & 0 & 9 & 1  \\\hline
6& 0 & 1 & 0 & 4 & 0 & 7 & 0 & 12 & 0 & 19 & 0 & 26 & 0 & 35 & 0 & 46 & 0 & 57 & 0  \\\hline
8& 0 & 1 & 1 & 5 & 1 & 14 & 2 & 30 & 0 & 55 & 3 & 91 & 1 & 140 & 4 & 204 & 0 & 285 & 5  \\\hline
10& 0 & 1 & 0 & 7 & 0 & 26 & 0 & 66 & 0 & 143 & 0 & 273 & 0 & 476 & 0 & 776 & 0 & 1197 & 0  \\\hline
12& 0 & 1 & 1 & 10 & 0 & 42 & 4 & 132 & 0 & 335 & 7 & 728 & 0 & 1428 & 12 & 2586 & 0 & 4389 & 19 \\\hline
14& 0 & 1 & 0 & 12 & 0 & 66 & 0 & 246 & 0 & 715 & 0 & 1768 & 0 & 3876 & 0 & 7752 & 0 & 14421 & 0  \\\hline
16& 0 & 1 & 1 & 15 & 1 & 99 & 5 & 429 & 1 & 1430 & 14 & 3978 & 2 & 9690 & 30 & 21318 & 0 & 43263 & 55  \\\hline
18& 0 & 1 & 0 & 19 & 0 & 143 & 0 & 715 & 0 & 2704 & 0 & 8398 & 0 & 22610 & 0 & 54484 & 0 & 120175 & 0  \\\hline
20& 0 & 1 & 1 & 22 & 0 & 201 & 7 & 1144 & 0 & 4862 & 26 & 16796 & 0 & 49742 & 66 & 130752 & 0 & 312455 & 143
   \\\hline
22& 0 & 1 & 0 & 26 & 0 & 273 & 0 & 1768 & 0 & 8398 & 0 & 32066 & 0 & 104006 & 0 & 297160 & 0 & 766935 & 0  \\\hline
24& 0 & 1 & 1 & 31 & 1 & 364 & 10 & 2652 & 0 & 14000 & 42 & 58786 & 4 & 208012 & 132 & 643856 & 0 & 1789515 & 335 \\\hline
26& 0 & 1 & 0 & 35 & 0 & 476 & 0 & 3876 & 0 & 22610 & 0 & 104006 & 0 & 400024 & 0 & 1337220 & 0 & 3991995 & 0  \\\hline
28& 0 & 1 & 1 & 40 & 0 & 612 & 12 & 5538 & 0 & 35530 & 66 & 178296 & 0 & 742900 & 246 & 2674440 & 0 & 8554275 & 715\\\hline
30& 0 & 1 & 0 & 46 & 0 & 776 & 0 & 7752 & 0 & 54484 & 0 & 297160 & 0 & 1337220 & 0 & 5170604 & 0 & 17678835 & 0
   \\\hline
32& 0 & 1 & 1 & 51 & 1 & 969 & 15 & 10659 & 1 & 81719 & 99 & 482885 & 5 & 2340135 & 429 & 9694845 & 1 & 35357670 & 1430
    \\\hline
34& 0 & 1 & 0 & 57 & 0 & 1197 & 0 & 14421 & 0 & 120175 & 0 & 766935 & 0 & 3991995 & 0 & 17678835 & 0 & 68635478 & 0
  \\\hline
36& 0 & 1 & 1 & 64 & 0 & 1463 & 19 & 19228 & 0 & 173593 & 143 & 1193010 & 0 & 6653325 & 715 & 31429068 & 0 & 129644790 &
   2704  \\\hline
38& 0 & 1 & 0 & 70 & 0 & 1771 & 0 & 25300 & 0 & 246675 & 0 & 1820910 & 0 & 10855425 & 0 & 54587280 & 0 & 238819350 & 0
    \\\hline
40& 0 & 1 & 1 & 77 & 1 & 2126 & 22 & 32890 & 0 & 345345 & 201 & 2731365 & 7 & 17368680 & 1144 & 92798380 & 0 & 429874830 &
   4862 \\\hline
      \end{tabular}\\[.5em]
       \label{tab:PFNs}
    \end{center}
   \caption{$N_{\rm forbidden}$, the number of Pauli-forbidden necklaces
        calculated from eq.~(\ref{PFNsBF})(entries with odd $F$ and/or odd
         $B$ vanish identically).}
 \end{table}
}
\end{landscape}

\end{document}